# Weak value amplification using *asymmetric* spectral response of Fano resonance as a natural pointer

Ankit K. Singh, Subir K. Ray, Shubham Chandel, Simonty Pal, Angad Gupta, P. Mitra* and N. Ghosh*

*Department of Physical Sciences,*

*Indian Institute of Science Education and Research (IISER) Kolkata.*

*Mohanpur 741246, India*

*Corresponding authors: pmitra@iiserkol.ac.in, nghosh@iiserkol.ac.in

**Abstract**

Weak measurement enables faithful amplification and high precision measurement of small physical parameters and is under intensive investigation as an effective tool in metrology and for addressing foundational questions in quantum mechanics. Most of the experimental reports on weak measurements till date have employed *external symmetric* Gaussian pointers. Here, we demonstrate its universal nature in a system involving *asymmetric* spectral response of Fano resonance as the pointer arising *naturally* in precisely designed metamaterials, namely, waveguided plasmonic crystals. The weak coupling arises due to a tiny shift in the *asymmetric* spectral response between two orthogonal linear polarizations. By choosing the pre- and post-selected polarization states to be nearly mutually orthogonal, we observe both real and imaginary weak value amplifications manifested as spectacular shift of the peak frequency of Fano resonance and narrowing (or broadening) of the resonance line width, respectively. Weak value amplification using *asymmetric* Fano spectral response broadens the domain of applicability of weak measurements using *natural* spectral line shapes as pointer in wide range of physical systems.

In 1988, Aharonov, Albert, and Vaidman introduced the concept of quantum weak measurements [1]. The proposed measurement process involves preparation of the system state in a definite initial state, which due to weak coupling to the observable results in a superposition of "slightly" shifted eigenstates and subsequent post selection in a final state which is nearly orthogonal to the initial state. The outcome of a weak measurement, which is called 'weak value', can become exceedingly large and lie outside the eigenvalue spectrum of the observable. This extraordinary feature of weak measurement has made it an attractive tool for addressing fundamental questions in quantum mechanics and for metrological applications [1-14]. This quantum mechanical concept can also be formulated within the realm of classical electromagnetic theory of light and indeed, carefully designed experiments were able to verify the validity of weak measurement in optical systems. For example, weak measurements have been employed to amplify tiny Spin Hall effect of light [3], Goos-Hänchen beam shift [15], to map the average trajectories of photons in double slit experiment [8], for the direct measurement of the quantum wave function of photon [9], for high resolution phase measurement [7], for sensitive estimation of angular rotations [4], and so forth.

Most of the experimental reports on weak measurements in the optical domain till date have employed *external symmetric* Gaussian pointers, e.g., Gaussian spatial modes of laser beams or Gaussian temporal pulse [1-5, 7-12]. Here, we demonstrate weak measurements using an *asymmetric natural* pointer generated within the system, that of the spectral response of Fano resonance in waveguided plasmonic crystals [16-18]. We have precisely fabricated plasmonic crystal samples, such that the *asymmetric* spectral response for two orthogonal linear polarizations are only slightly shifted (unresolved), thus mimicking the basic criterion for weak measurement. Pre and post selections in prescribed polarization states lead to weak value amplification, manifested as spectacular changes in the asymmetric line shape of Fano resonance. The shift in the peak frequency and the change in the spectral width of the resonance are attributed to real and the imaginary weak-value amplification, respectively. The results are shown to match with appropriate theoretical model. This is the first demonstration of weak measurements using an *asymmetric natural* pointer that of the spectral response of Fano resonance, which should broaden the horizon of weak measurements and open up new avenues for the use of such *natural* spectral line shapes as pointers for weak measurement in wide class of spectroscopic systems.

The asymmetric spectral line shape in Fano resonance arises due to the interference of a narrow resonance with a broad spectrum [16-25]. The waveguided plasmonic crystal is an interesting nano optical system that exhibits prominent Fano resonance [16-18], and has been studied for numerous potential applications [17, 18]. Such systems usually comprise of a periodic array of noble metal nanostructures on top of a dielectric waveguiding layer [16-18]. The coupling of the surface plasmons in the metallic nanostructures and the quasi-guided photonic modes in the waveguiding layer leads to Fano resonance [16-18]. The electric field of the scattered light from such system can be modeled as the interference of a narrow resonance, whose field amplitude is described by a complex Lorentzian $L^R(\omega)$, with a broad spectrum having relative amplitude $B$ [18]

$$E_s(\omega) = [L^R(\omega) + B] = \left[\frac{q-i}{\varepsilon(\omega)+i} + B\right] \tag{1}$$

Here, $\varepsilon(\omega) = \frac{\omega-\omega_0}{(\gamma/2)}$, $\omega_0$ and $\gamma$ are the central frequency and the width of the narrow resonance, respectively. The Fano asymmetry parameter $q$ depends upon the strength of coupling of the two modes and determines the asymmetry of the line shape [16-19]. The intensity corresponding to Eq. 1 ($I_s(\omega) = |E_s(\omega)|^2$) comprises of a Fano resonant term exhibiting the asymmetric spectral line shape and a Lorentzian background [18] (Supporting information, **Eq. S1**) and the frequencies corresponding to the minimum and the maximum of Fano resonance are

$$\omega_F = \left(\omega_0 - \frac{q\gamma}{2}\right); \quad \omega_m = \left(\omega_0 + \frac{\gamma}{2q}\right) \tag{2}$$

Depending upon the geometry and optical properties of a given Fano resonant system, different spectral response for orthogonal linear polarizations [16-18] can be observed, which can be modeled using Jones matrix ($J(\omega)$) [26]

$$\boldsymbol{E_s}(\omega) = J(\omega)|\psi_{in}\rangle; \quad J(\omega) = \begin{pmatrix} j_x & 0 \\ 0 & j_y \end{pmatrix}; \quad j_{x/y} = L^R_{x/y}(\omega) + B_{x/y}; \quad L^R_{x/y} = \frac{q_{x/y}-i}{\varepsilon_{x/y}+i} \tag{3}$$

Here, the polarization state of input light is represented by the Jones vector $|\psi_{in}\rangle$ and $\boldsymbol{E_s}$ is the scatted electric field vector. The geometrical structure of plasmonic crystals can be tailored so that the central frequencies of the narrow resonance corresponding to the transverse magnetic (TM- $x$) and the transverse electric (TE- $y$) linear polarizations are slightly shifted (unresolved) with respect to each other ($\omega_{ox} = \omega_o + a$, $\omega_{oy} = \omega_o - a$, $2a \ll \gamma$) [16], which can be considered as a manifestation of weak interaction between the polarization and the spectral response. For successful weak measurement using the asymmetric Fano spectral line shape as pointer, one would simultaneously need to fulfill the criterion $(\omega_{mx} - \omega_{my} \ll \gamma)$ and $(\omega_{Fx} - \omega_{Fy} \ll \gamma)$. This implies that the difference in the $q$-parameters between the $x$ and $y$ polarizations has to be small $(q_x \sim q_y)$. The corresponding criteria for weak measurements using Fano resonance can be put forward as

$$(q_x - q_y) \ll 2, \ (q_{x/y} > 1); \ \left(\frac{1}{q_x} - \frac{1}{q_y}\right) \ll 2, \ (q_{x/y} < 1) \tag{4}$$

In what follows, we describe two schemes for the real and the imaginary weak value amplification of the polarization operator using asymmetric Fano spectral line shape as pointer. We consider the simpler case corresponding to $q_x = q_y = q$, $B_x = B_y = 1$ in Eq. 3.

*Scheme 1: Real weak value amplification*

The input state $|\psi_{in}\rangle$ is taken to be $+45^0$ (diagonal) linear polarization and post selections ($|\psi_{post}\rangle$) are done at nearly orthogonal linear polarizations as given by

$$|\psi_{in}\rangle = [1, \ 1]^T; \ |\psi_{post}\rangle = \begin{bmatrix} \cos(\epsilon) \pm \sin(\epsilon) \\ -\cos(\epsilon) \pm \sin(\epsilon) \end{bmatrix}^T$$

Here, $\epsilon$ is a small angle by which the post selected state is away from orthogonal to the input state. The resultant electric field be obtained as $E_s(\omega) = \langle\psi_{post}|J(\omega)|\psi_{in}\rangle$ and the corresponding intensity ($I_s(\omega) = |E_s(\omega)|^2$) gives rise to asymmetric Fano spectral line shape as shown in **Figure 1a**. For small $a$ and $\varepsilon$, $E_s(\omega)$ can be approximated as

$$E_s(\omega) \approx 2\sin(\epsilon)\left[\frac{(q-i)\gamma/2}{\{\omega-(\omega_o \pm a\cot\epsilon)+i\gamma/2\}} + 1\right] \tag{5}$$

Equation 5 clearly shows that as $\varepsilon$ becomes small a large frequency shift in the peaks of the contributing narrow resonance ($\delta\omega_0 = \pm a\cot\epsilon$) (**Figure 1b**) and the resulting Fano resonance ($\delta\omega_m = \pm a\cot\epsilon$) (**Figure 1c**) occurs. These large frequency shifts can also be modeled as the amplification of the real weak value ($A_W$) of the $2 \times 2$ diagonal polarization operator $\boldsymbol{\sigma}$, which has eigenvalues $\pm a$ [2].

$$A_W = \frac{\langle\psi_{post}|\sigma|\psi_{pre}\rangle}{\langle\psi_{post}|\psi_{pre}\rangle} = \pm a\cot\epsilon \tag{6}$$

Note, in this case, $|\psi_{pre}\rangle = |\psi_{in}\rangle$.

*Scheme 2: Imaginary weak value amplification*

Here, post-selections are done at nearly orthogonal elliptical polarizations, given by

$$|\psi_{in}\rangle = [1,\ 1]^T;\ |\psi_{post}\rangle = [1, -e^{\pm i\epsilon}]^T$$

where $\epsilon$ represents small phase shifts with respect to the orthogonal state. The electric field of light can be obtained as $E_s(\omega) = \langle\psi_{post}|J(\omega)|\psi_{in}\rangle$, which gives rise to asymmetric Fano spectral line shapes (**Figure 1d**). Again, for small $a$ and $\epsilon$, $E_s(\omega)$ can be approximated as

$$E_s(\omega) \approx 2i\sin\left(\frac{\epsilon}{2}\right)\left[\frac{(q-i)\frac{\gamma}{2}}{\{\omega-\omega_o+i\left(\frac{\gamma}{2}\pm a\cot(\epsilon/2)\right)\}} + 1\right] \quad (7)$$

Unlike scheme 1, here the spectral width of the resonance is considerably broadened or narrowed ($\delta\gamma = \pm 2a\cot(\epsilon/2)$) (**Figure 1e**). Moreover, this modification in $\gamma$ also leads to a shift in the frequency peak of the Fano resonance ($\delta\omega_m = \pm\frac{2a}{q}\cot(\epsilon/2)$) (Eq. 2) (**Figure 1f**). The broadening (narrowing) of the spectral width of the resonance can be modeled as the amplification of the imaginary weak value ($A_W$) of $\boldsymbol{\sigma}$ as [2]

$$A_\omega = \frac{\langle\psi_{post}|\sigma|\psi_{pre}\rangle}{\langle\psi_{post}|\psi_{pre}\rangle} \approx \pm ia\cot(\epsilon/2) \quad (8)$$

The above schemes are also applicable for small differences in the parameters ($q_x \neq q_y$, $B_x \neq B_y$) which is the case for our experimental situations. For such cases the pre-selected state is not the input state, but rather $|\psi_{pre}\rangle = \boldsymbol{R}|\psi_{in}\rangle$. Here, $\boldsymbol{R}$ is a strong operator, defined by the Jones matrix (Eq. 3) at $\omega = \omega_0$, which encodes anisotropy of the system (due to $q_x \neq q_y$, $B_x \neq B_y$). The weak measurement schemes may then be implemented by choosing an elliptical input polarization state to account for the anisotropy effect so that $|\psi_{pre}\rangle = [1,\ 1]^T$ (see Supporting information **Figure S1**).

The schematics of our experiments are shown in **Figure 2a**. The heart of our experiment is a custom designed optical setup where one can determine the complete polarization response of a sample by recording the $4 \times 4$ spectral Mueller matrix $M(\omega)$ with great precision (see Supporting information). This home-built spectroscopic Mueller matrix system integrated with a dark field microscope employs broadband white light excitation and subsequent recording of sixteen polarization resolved scattering spectra (wavelength $\lambda = 400 - 725$ nm, 1 nm resolution) by sequential generation and analysis of four optimized elliptical polarization states (see Supporting information, **Figure S2**). This elegant technique enables us to extract the spectral response of a given sample for any well-defined polarization directions of the input and the scattered light. We fabricated waveguided plasmonic crystal samples comprising of one dimensional (1-D) periodic gold (Au) grating on top of indium tin oxide (ITO) waveguiding layer. The fabrication of the nanostructure involved Electron beam lithography and metal deposition by thermal evaporation technique (see Supporting information). The thickness of the waveguiding ITO layer is 190nm. The dimension of the fabricated Au grating was (width = 83 nm, height = 20 nm, centre to centre distance = 550 nm). These geometrical parameters of the grating were chosen so that the asymmetric spectral response for the two orthogonal linear polarizations (TM- $x$ and TE- $y$) are only slightly shifted (unresolved), thus mimicking the basic criterion for weak measurement. The overall dimension of grating structure was $300 \times 300 \ \mu m^2$ and the spot size of the beam at the sample site was ~3 mm. Typical SEM image of Au grating is shown in **Figure 2b**. The weak measurement schemes were implemented by performing pre and the post selections of the polarization states on $M(\omega)$

$$I_s(\omega) = |\langle \psi_{post}|J(\omega)|\psi_{in}\rangle|^2 = \frac{1}{2} S_{post}^T M(\omega) S_{in} \qquad (9)$$

Here, $S_{post}$, $S_{in}$ are the Stokes vectors [26] corresponding to the Jones vectors $|\psi_{post}\rangle$ and $|\psi_{in}\rangle$. Typical polarization blind ($M_{11}$ Mueller matrix element) scattering spectra of the plasmonic crystal sample exhibits characteristically asymmetric spectral line shape (**Figure 2c**). As noted previously, the interference of the scattered fields of the surface plasmon resonance of the Au grating (acting as the broad continuum) with the quasi-guided modes of the underlying ITO waveguide layer (acting as the narrow resonance peaking at $E_0 = \hbar\omega_0$ ~1.803 $eV$) leads to Fano resonance ($E_m = \hbar\omega_m$~ 1.837 eV).

The spectral Mueller matrix $M(\omega)$ of the Au grating (**Figure 3a**) exhibits nearly block diagonal structure ($M_{34}/M_{43} \gg M_{24}/M_{42}$, and $M_{12}/M_{21} \gg M_{13}/M_{31}$), confirming the laboratory $x$ and $y$ linear polarizations as the system eigenstates. The spectral line shapes for the orthogonal $x$ and $y$ linear polarizations (**Fig. 3b and Fig. 3c**) appear overlapping (unresolved) and the estimates for the resonance parameters also exhibit small differences ($E_{0x} - E_{0y} \ll \gamma$, $q_x \sim q_y$, $B_x \sim B_y$), satisfying the criteria for weak measurements. Note that even though the periodicity of the Au grating can be tailored to ensure that the peak of the Fano resonance nearly overlaps for the TM($x$) and transverse electric TE ($y$) linear polarizations for plane wave excitation, the overall spectral line shapes may still be different. However, unlike plane wave excitation, in our high NA microscopic geometry, a significant fraction of the scattered light is due to excitation with scrambled polarization even when the input states are either $x$ or $y$ polarized. This turns out to be beneficial for mimicking the basic criterion for weak measurement, as evident from the nearly overlapping spectral line shapes corresponding to the $x$ and $y$ polarizations (**Fig. 3b and Fig. 3c**).

**Figure 4** presents the results of weak measurements using *scheme 1* (Fig. 4a, 4b) and *scheme 2* (Fig. 4c, 4d). Pre and the post selections of the polarization states were performed on the recorded Mueller matrix $M(\omega)$ to realize the weak measurement schemes. The input state ($S_{in} = [1, -0.0493, 0.9833, -0.1752]^T$) was chosen to be elliptical instead of $+45^0$ linear polarization in order to account for the non-zero phase and amplitude anisotropy ($q_x \neq q_y$, $B_x \neq B_y$ in Fig. 3b and 3c) of the plasmonic crystal and the wavelength independent background phase anisotropy of the system (linear retardance $L_R \sim 0.25\ rad$, see **Figure S3** of Supporting information). With this choice of $S_{in}$, the pre-selected state became $+45^0$ linear polarization ($S_{pre} = [1, 0, 1, 0]^T$ corresponding to $|\psi_{pre}\rangle = [1,\ 1]^T$) as desired for the implementation of our weak measurement schemes. This was verified further by varying the post selected state ($S_{post}$) on $M$ and by identifying the state for which the overall intensity was minimum. This post selected state was exactly orthogonal to the pre-selected state, which in our case was $S \approx [1, 0, -1, 0]^T$. The weak measurements were then performed around this orthogonal state: (a) $\pm 2\epsilon$ rotation of $S$ for implementing *scheme 1* ($S_{post} = [1, \pm sin2\epsilon, \mp cos2\epsilon, 0]^T$), (b) phase retardation of $\pm \epsilon$ for implementing *scheme 2* ($S_{post} = [1, 0, \pm cos\epsilon, \mp sin\epsilon]^T$). The corresponding results are summarized in (Fig. 4a, 4b), and (Fig. 4c, 4d), respectively. Post selection using *scheme 1* leads to rapid shift of the Fano resonance peak ($E_m$) with varying $\epsilon$ ($= 0.5\ to\ 0.05\ rad$) (**Fig. 4a**). The shift in the peak of the narrow waveguide resonance ($\delta E_0$) and the Fano resonance ($\delta E_m$) (**Fig. 4b**) appear to scale as $\sim cot\epsilon$, demonstrating faithful amplification of the real weak value. For post selection using *scheme 2*, the spectral width $\gamma$ of the resonance increases with decreasing phase angle $\epsilon$ (**Fig. 4c**), as a manifestation of the imaginary weak value amplification. The corresponding change in $\delta\gamma$ (**Fig. 4d**) also shows qualitative agreement with the prediction of Eq. 8 ($\sim cot(\epsilon/2)$). However, shift in the Fano resonance peak ($\delta E_m$ in *scheme 1*) and resonance width $\delta\gamma$ (in *scheme 2*) show slight deviation from ideal weak value amplification. The difference in the $q$-parameters between the TM ($x$) and TE ($y$) polarizations ($q_x \neq q_y$) is primarily responsible for this deviation (see Supporting information **Figure S1**). Nevertheless, these results provide conclusive evidence of both real and imaginary weak value amplification using the Fano spectral response as a *natural* pointer.

We have demonstrated a new concept of weak measurement using the *asymmetric* spectral response of Fano resonance as the *natural* pointer. This offered remarkable control over both the asymmetric line shape and the line width of Fano resonance, which may significantly impact Fano resonance-based applications in the nano optical domain [17, 27] and beyond [24, 28], where precision control of the Fano line shape is desirable. We believe that our protocol opens up new possibilities of weak value amplification for high precision measurement of small physical quantities [29, 30] or for enhanced probing of interactions [19, 20, 22] in a wide array of physical systems where Fano resonance is used as a fingerprint [29]. In general, weak value amplification using asymmetric Fano spectral response as pointer opens up new paradigm of weak measurements using *natural* spectral line shapes in wide class of spectroscopic systems [6, 19, 20, 22, 30].

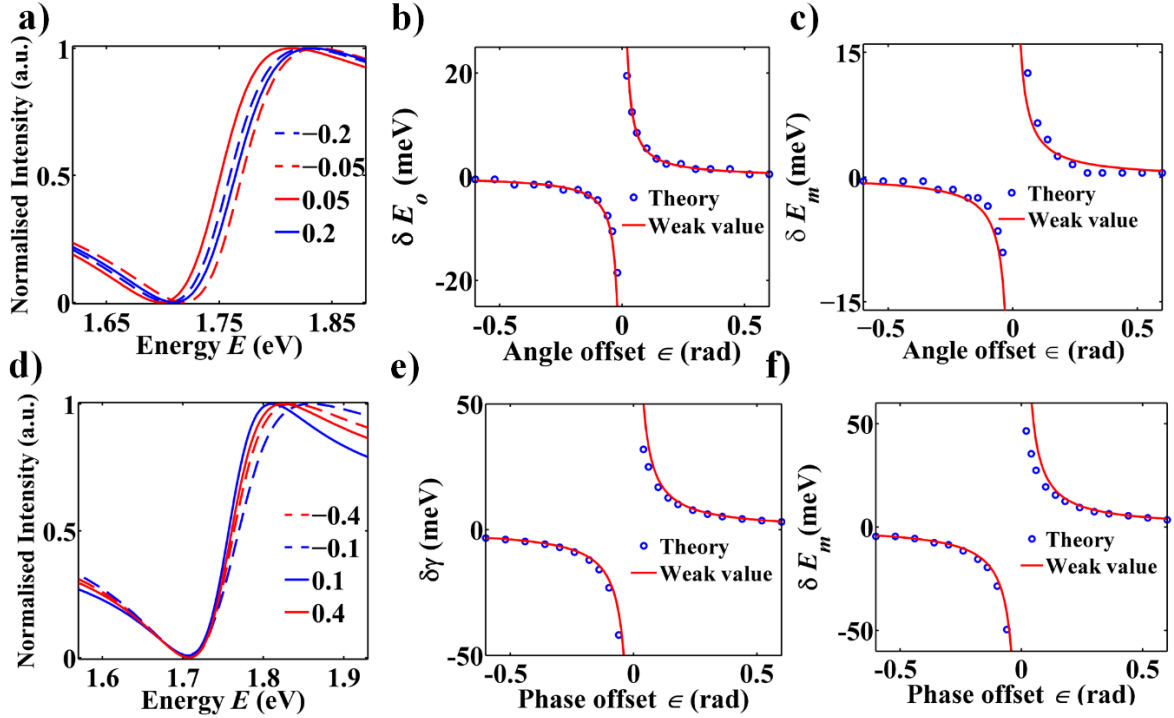

**Figure 1: Theoretical simulation of weak measurements using asymmetric Fano spectral line shape for *scheme 1* (a, b and c) and *scheme 2* (d, e and f).** The spectral variation of the intensity were simulated using pre and post selection of the polarization states on the Jones matrix of Eq. 3. The input parameters were $E_{0x} = 1.756\ eV$, $E_{0y} = 1.755\ eV$, width $\gamma_x = \gamma_y = 0.12\ eV$, $q_x = q_y = 0.8$. **(a)** The resulting spectra for post selections with two different small angles $\pm\epsilon$ using *scheme 1*. The shift in the peaks of **(b)** the contributing narrow resonance ($\delta E_0$) and **(c)** the Fano resonance ($\delta E_m$) with varying $\epsilon$ (shown by symbols, open circles). The corresponding fits with Eq. 6 ($\sim \pm \cot\epsilon$) for real weak value amplification are shown by solid lines. **(d)** The spectra for post selections with two different small phase offset $\pm\epsilon$ using *scheme 2*. **(e)** The changes in the spectral width ($\delta\gamma$) and **(f)** the shift of the Fano resonance peak ($E_m$) with varying $\epsilon$ are shown by symbols (open circles). The corresponding fits with Eq. 8 ($\sim \pm \cot(\epsilon/2)$) for imaginary weak value amplification are shown by solid lines.

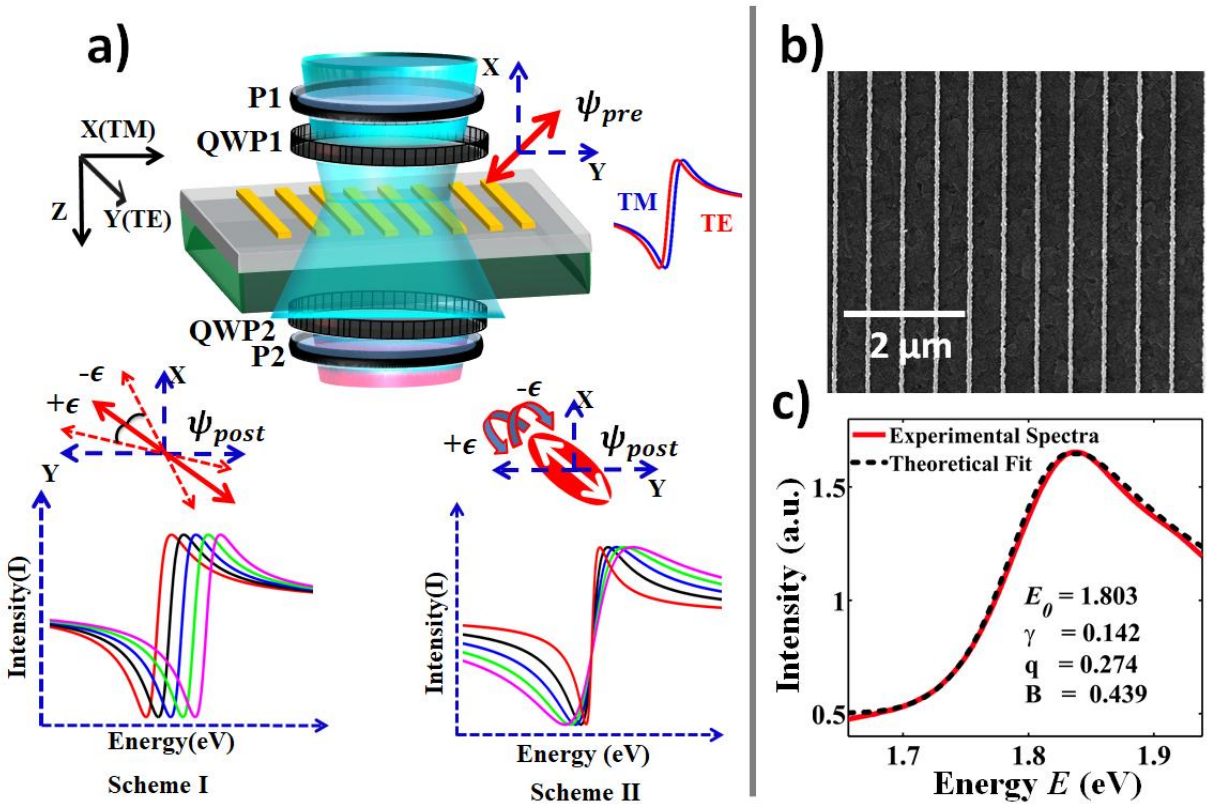

**Figure 2: Weak measurements using the asymmetric Fano spectral response of waveguided plasmonic crystal as a natural pointer.** (a) Schematic illustration of the concept. A tiny shift in the Fano spectral response of the plasmonic crystal between the transverse magnetic (TM-*x*) and the transverse electric (TE-*y*) polarizations provides the weak coupling. The pre-selected state is chosen to be of ($+45^0$) linear polarization. The plasmonic crystal generates a tiny shift in the Fano spectral response for the transverse magnetic (TM-x) and the transverse electric (TE-y) polarizations. Post selections in $\pm \epsilon$ small angle away from the orthogonal linear polarization ($+135^o$) state lead to real weak value amplification, manifesting as a shift of the Fano resonance peak (*Scheme 1*). Post selections in nearly orthogonal elliptical polarization states ($\pm \epsilon$ small phase offset from the orthogonal state) lead to imaginary weak value amplification, manifesting as narrowing (or broadening) of the resonance line width. **(b)** Typical SEM images of Au grating (1-D plasmonic crystal). The dimension of the Au grating: width = 83 nm, height = 20 nm, centre to centre distance = 550 nm. **(c)** The scattering spectra ($E = 1.650$ to $1.940$ eV, corresponding $\lambda \approx 750 - 640$ nm, shown here) from Au grating (red solid line) showing typical Fano spectral asymmetry. Theoretical fit of the spectra with Fano intensity expression (Supporting information Eq. S1) is shown by black dotted line. The values for the fitted parameters of resonance are $E_0 = 1.803\ eV$, $E_m = 1.837$ eV, width $\gamma = 0.142$, $q = 0.274$, $B = 0.439$.

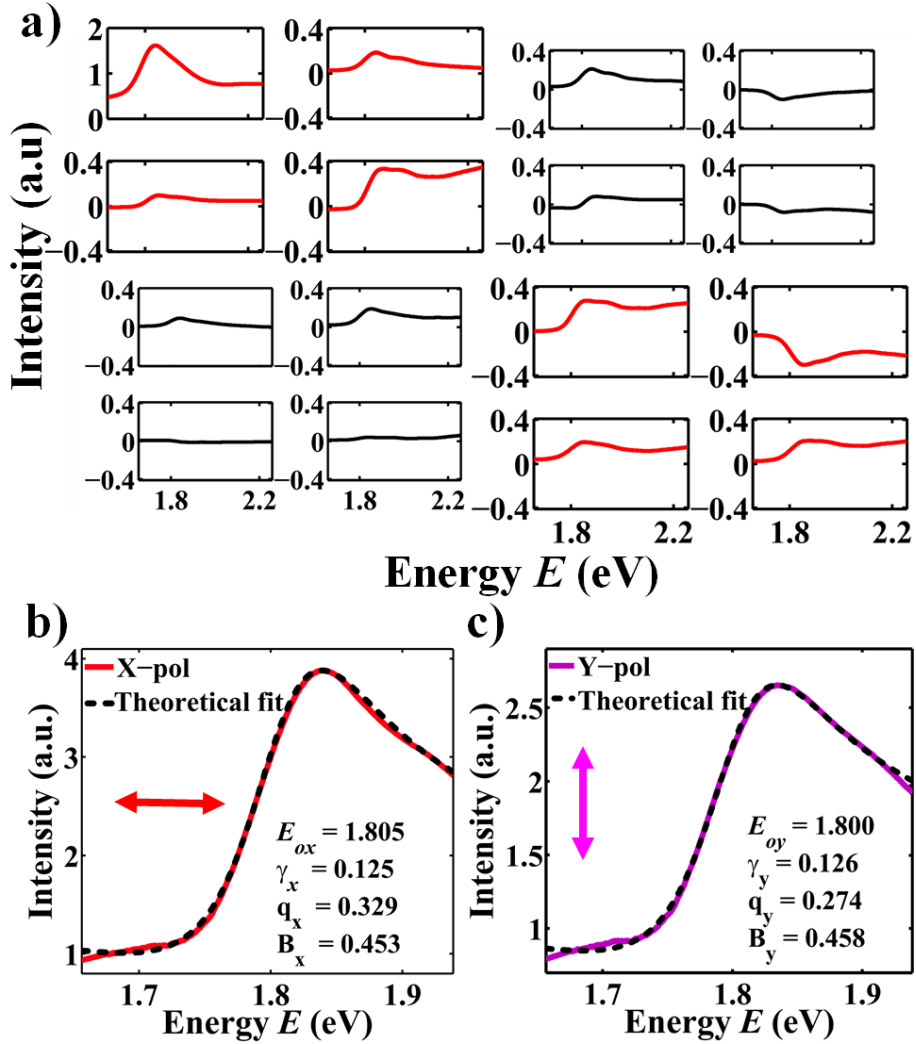

**Figure 3: Confirmation of weak coupling between the polarization and the Fano spectral response of plasmonic crystal from experimental Mueller matrix.** (**a**) The scattering spectral Mueller matrix $M(E)$ (for $E = 1.650$ to $2.250$ eV) for the fabricated 1-D plasmonic crystal of Au grating. The spectral variation of scattered intensities ($E = 1.650$ to $1.940$ eV) for (**b**) $x$-polarization (red solid line) and (**c**) $y$-polarization (magenta solid line). The spectra were obtained by pre and post selection (using Eq. 9) of the corresponding polarization states ($\boldsymbol{S}$: $[1, 1, 0, 0]^T$ for $x$ and $[1, -1, 0, 0]^T$ for $y$) on $M$. Theoretical fits of the spectra with Fano intensity expression (Supporting information Eq. S1) are shown by black dotted lines. The values for the fitted parameters of resonance for the two orthogonal linear polarizations are $E_{0x} = 1.805\ eV$, $E_{0y} = 1.800\ eV$, width $\gamma_x = 0.125\ eV, \gamma_y = 0.126\ eV$, $q_x = 0.329$, $q_y = 0.274$, confirming the validity of weak measurements criterion (Eq. 4).

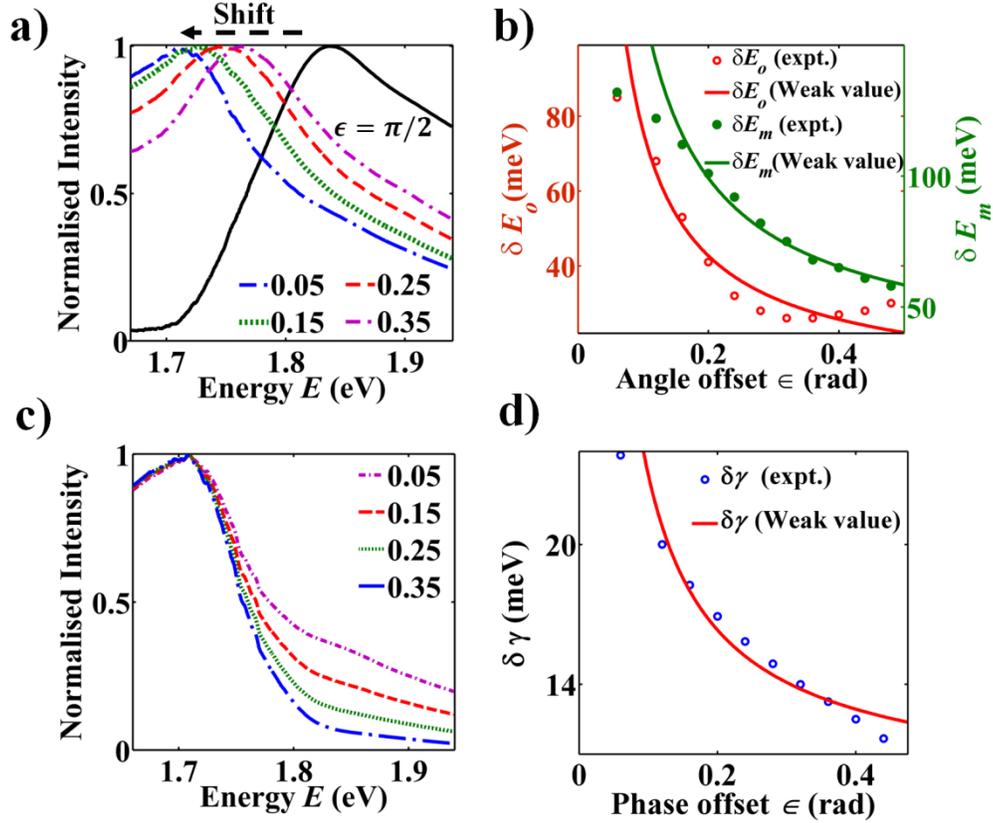

**Figure 4: Experimental demonstration of weak measurements using asymmetric Fano spectral line shape in plasmonic crystal (corresponding to Fig. 3).** Results of weak measurements using *scheme 1* shown in (a) and (b), and that for *scheme 2* shown in (c) and (d). **(a)** The resulting spectra for post selections with four different small angles $\epsilon$ using *scheme 1*. The spectra for a *strong* measurement corresponding to pre and post selections at $+45^0$ linear polarization (black line) is shown as a reference. The shift in (b) the peak of the narrow waveguide resonance ($\delta E_0$, left axis, red) and the peak of Fano resonance ($\delta E_m$ right axis, green) with varying $\epsilon$ are shown by symbols (circles). The corresponding fits with Eq. 6 ($\sim cot\epsilon$) for real weak value amplification are shown by solid lines. **(c)** The spectra for post selections with four different small phase offset $\epsilon$ using *scheme 2*. **(d)** The changes in the spectral width ($\delta\gamma$) with varying $\epsilon$ is shown by symbols (open circles). The corresponding fits with Eq. 8 ($\sim \pm \cot(\epsilon/2)$) for imaginary weak value amplification is shown by solid lines.

# Supporting Information

# Weak value amplification using *asymmetric* spectral response of Fano resonance as a natural pointer

Ankit K. Singh, Subir K. Ray, Shubham Chandel, Semanti Pal, Angad Gupta, Partha Mitra* and Nirmalya Ghosh*

*Department of Physical Sciences,*

*Indian Institute of Science Education and Research (IISER) Kolkata.*

*Mohanpur 741246, India*

*Corresponding authors: pmitra@iiserkol.ac.in, nghosh@iiserkol.ac.in

1. **Theoretical treatment on weak measurements using asymmetric spectral response of Fano resonance as pointer**

**A.** *Expression for intensity of Fano resonance in scattering*

The expression for the scattered intensity corresponding to the electric field of Eq. 1 of the manuscript is

$$I_s(\omega) = |E_s(\omega)|^2 = B^2 \left[\frac{(q/B+\varepsilon)^2}{\varepsilon^2+1}\right] + \frac{(B-1)^2}{\varepsilon^2+1} \quad \text{(S1)}$$

The first term represents the Fano-type asymmetric spectral line shape with an effective asymmetry parameter $q^{eff} = q/B$. The second term corresponds to a Lorentzian background, widely reported in the context of Fano resonance on diverse optical systems [1]. The intensity data ($I_s(\omega)$) following each pre and post selections were fitted with Eq. S1 to estimate the resonance parameters (peak of the contributing narrow resonance $E_0 = \hbar\omega_0$, peak of the Fano resonance $E_m = \hbar\omega_m$, and the width of resonance $\gamma$).

**B.** *Weak measurement scheme 1 and scheme 2 for $q_x \neq q_y$*

In the main text of the manuscript, weak measurement scheme 1 and scheme 2 are presented for the case when the Fano asymmetry parameter is the same for orthogonal *x* and *y* linear polarizations ($q_x = q_y = q$). The schemes are also applicable for small difference in the $q$ parameters ($q_x \neq q_y$, satisfying Eq. 4). In such case, the pre-selected state ($\psi_{pre}$) is not the input polarization state ($\psi_{in}$), rather the state will be $|\psi_{pre}\rangle = R|\psi_{in}\rangle$. Here, $R$ is a strong operator, defined by the Jones matrix (Eq. 3) at $\omega = \omega_0$, which encodes the phase and the amplitude anisotropy of the system (due to $q_x \neq q_y$, $B_x \neq B_y$).

$$R = \begin{pmatrix} \alpha_x & 0 \\ 0 & \alpha_y \end{pmatrix}; \quad \alpha_x = (B_x - 1) - iq_x; \quad \alpha_y = (B_y - 1) - iq_y \quad \text{(S2)}$$

The *scheme 1 and 2* of weak measurements may then be implemented by choosing the input polarization state $|\psi_{in}\rangle = [\frac{1}{\alpha_x}, \frac{1}{\alpha_y}]^T$ so that $|\psi_{pre}\rangle = [1, 1]^T$. In general, any additional background (frequency-independent) phase and amplitude anisotropy of the system can also be taken into account in a similar manner while deciding the input polarization state (which will be elliptical in general) for implementing the weak measurement schemes.

In Figure 1 of the manuscript, theoretical results of weak measurements using *scheme 1* and *scheme 2* were presented for the case ($q_x = q_y = q$). Here, in **Figure S1**, we present the corresponding theoretical results for the case $q_x \neq q_y$, satisfying the criteria for weak measurement using the asymmetric spectral line shape of Fano resonance as pointer (Eq. 4). For this purpose, the intensities ($I_s(\omega)$) are simulated using pre and post selections of polarization states (according to scheme 1 and 2) in Eq. (3) and Eq. (9). Like in Figure 1, here also, real weak value amplification (*scheme 1*) is manifested shift in the peak frequencies of the contributing narrow resonance ($\delta E_0$) and the resulting Fano resonance ($\delta E_m$). Similarly, post selections using *scheme 2* lead to lead to significant broadening or narrowing (for $\pm \epsilon$) of the spectral width $\gamma$ of the resonance. However, shift in the peak frequency of Fano resonance ($\delta E_m$ in *scheme 1*) and changes in the resonance width $\delta\gamma$ (in *scheme 2*) with varying small post selection parameter $\epsilon$ are observed to deviate slightly from ideal behaviour (predictions of Eq. 6 and 8, respectively). Note that while the resulting effective Fano asymmetry parameter ($q$) does not change with varying post selections for the case of $q_x = q_y$, it varies with post selections for the case $q_x \neq q_y$. This is one of the reasons for the observed deviation from ideal weak value amplification. In general, with increasing difference $q_x \neq q_y$, the exact expression for the electric field ($E_s(\omega)$) following pre and post selections starts deviating from the approximate expressions of Eq. 5 (weak value amplification using *scheme 1*) and Eq. 7 (weak value amplification using *scheme 2*).

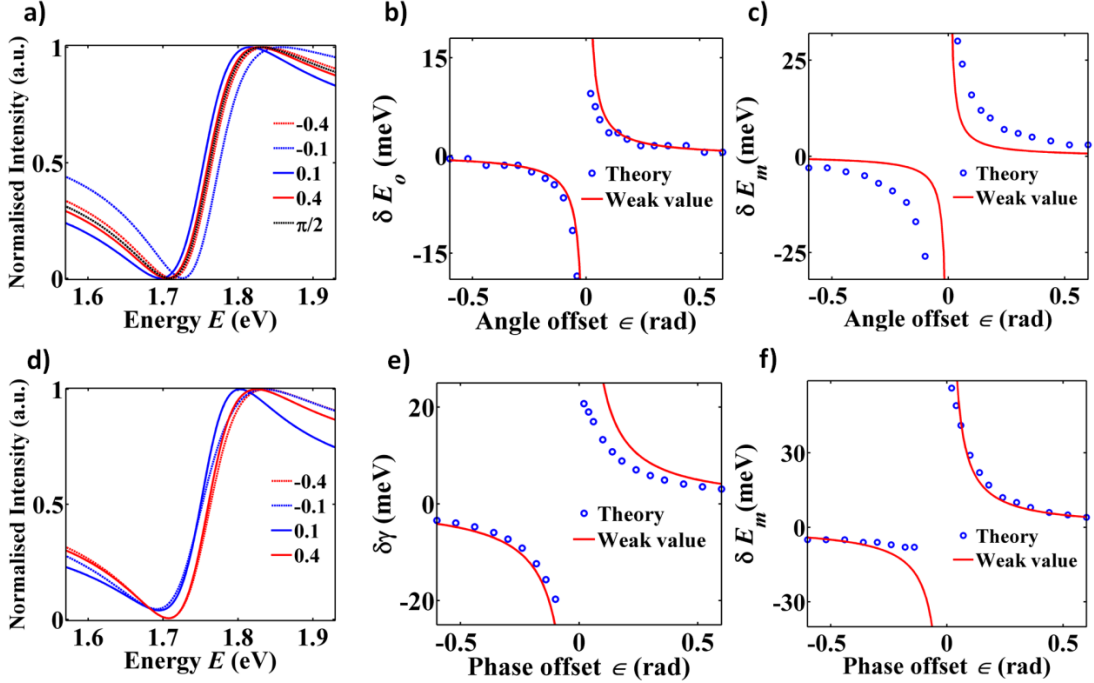

**Figure S1: Theoretical demonstration of the validity of weak measurements using asymmetric Fano spectral line shape.** Results of weak measurements using *scheme 1* (a, b and c) and *scheme 2* (d, e and f). The spectra ($I_s(\omega)$) were simulated using Eq. 3 and Eq. 9. The input parameters were ($E_{0x} = 1.756\ eV$, $E_{0y} = 1.755\ eV$, width $\gamma_x = \gamma_y = 0.12\ eV$, $q_x = 0.77, q_y = 0.8$). **(a)** The resulting spectra for post selections with two different small angles $\pm\epsilon$ using *scheme 1*. The systematic shift of the Fano resonance peak ($E_m$) with varying $\epsilon$ is highlighted in the inset. The shift in the peaks of the **(b)** contributing narrow resonance ($\delta E_0$) and **(c)** the Fano resonance ($\delta E_m$) with varying $\epsilon$ are shown by symbols (open circles). The corresponding fits with Eq. 6 ($\sim \pm cot\epsilon$) for real weak value amplification are shown by solid lines in the figures. **(d)** The spectra ($I_s(\omega)$) for post selections with two different small phase offset $\pm\epsilon$ using *scheme 2*. **(e)** The changes in the spectral width ($\delta\gamma$) and (f) the shift of the Fano resonance peak ($E_m$) with varying $\epsilon$ are shown by symbols (open circles). The corresponding fits with Eq. 8 ($\sim \pm \cot(\epsilon/2)$) for imaginary weak value amplification are shown by solid lines in the figures.

2. **Dark field Mueller matrix spectroscopy system**

The complete polarization response of the samples were determined by recording the $4 \times 4$ elastic scattering spectral Mueller matrices $M(\omega)$ [2] using a home-built spectroscopic Mueller matrix system integrated with a dark field microscope (IX71, Olympus) (**Figure S2**) [3]. Collimated white light from a halogen lamp (JC12 V100WHAL-L, Olympus) is used as an excitation source and is passed through the polarization state generator (PSG) unit for generating the input polarization states. The PSG unit consists of a horizontally oriented fixed linear polarizer P1 and a rotatable achromatic quarter waveplate (QWP1, AQWP05M-600, Thorlabs, USA) mounted on a computer controlled rotational mount (PRM1/M-Z7E, Thorlabs, USA). The light is then focused to an annular shape at the sample site using a dark-field condenser (Olympus U-DCD, NA =0.92). The sample-scattered light is collected by the microscope objective (MPlanFL N, NA =0.8), passed through the polarization state analyser (PSA) unit and is then relayed to spectrometer (USB4000, Ocean optics, USA) for spectrally resolved signal detection (scattering spectra. The PSA unit essentially comprises of the same components with a fixed linear polarizer (P2, oriented at vertical position) and a computer controlled rotating achromatic quarter waveplate (QWP2), but positioned in a reverse order.

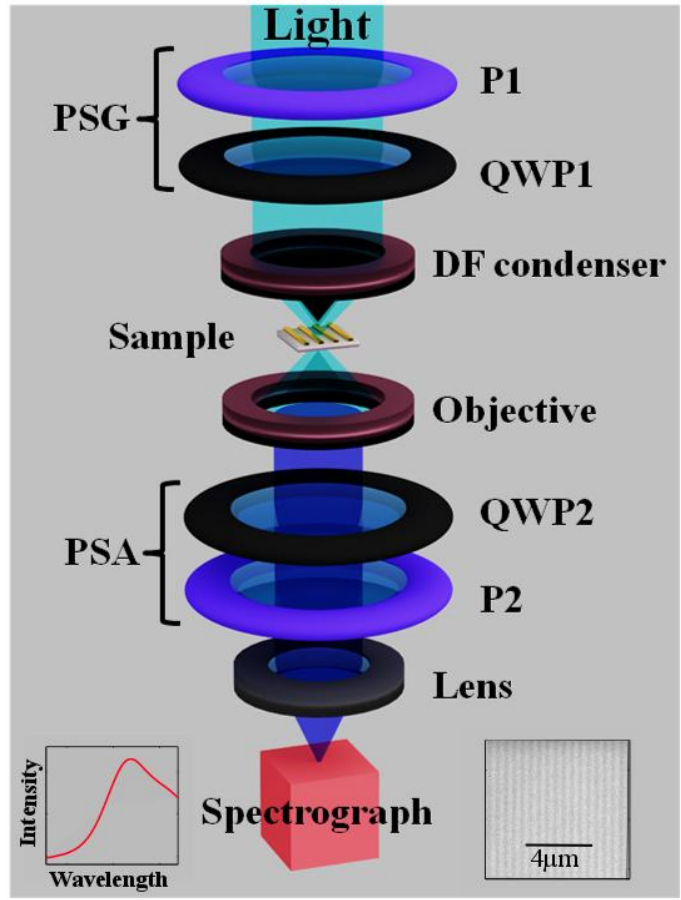

**Figure S2:** A schematic of the Dark-field Mueller matrix spectroscopic microscopy experimental system. PSG: Polarization state generator, PSA: Polarization state analyzer. (P1, P2): fixed linear polarizers, (QWP1, QWP2): rotatable achromatic quarter waveplates, DF: dark field condenser. (Inset left): Typical spectral response from a waveguided plasmonic crystal. ( Inset right): Dark field optical image of a grating sample.

The details of the Mueller matrix measurement strategy can be found elsewhere [2-6].The 4×4 spectral Mueller matrices were constructed by combining sixteen spectrally resolved intensity measurements (spectra) for four different combinations of the optimized elliptical polarization state generator (using the PSG unit) and analyzer (using the PSA unit) basis states. The fast axis of motorized quarter wave plate QWP1 is sequentially changed to four angles (35º, 70º, 105º and 140º) with respect to the axis of P1. These four sets of Stokes vectors (4×1 vector) are grouped as column vectors to form the 4×4 generator matrix *W*. Similarly the elliptical analyzer basis state is acquired by rotating the QWP2 fast axis to the corresponding four angles (35º, 70º, 105º and 140º). The analyzer states are analogously written as a 4×4 analyzer matrix $A$. The sixteen sequential intensity measurements (at any wavelength) are grouped in a 4×4 matrix $M_i$, which is related to $A$, $W$ matrices and the sample Mueller matrix $M$ as [2-5]

$$M_i = AMW$$

The Mueller matrix M can be determined using known forms of the $W$ and $A$ matrices as

$$M = A^{-1} M_i W^{-1}$$

Thus, the recorded sixteen polarization-resolved scattering spectra are pooled to yield the scattering Mueller matrix of the sample. The exact experimental forms of the $W(\lambda)$ and $A(\lambda)$ matrices and their wavelength dependence were determined using Eigen Value calibration method [2-5]. This approach facilitated recording of the spectral polarization response (Mueller matrix) exclusively of the plasmonic or other scattering samples, ensuring high accuracy of Mueller matrix measurement over a broad wavelength range [2,3].

### 3. Determination of the sample anisotropy parameters from Mueller matrix

The experimental Mueller matrix $M$ was decomposed using the polar decomposition approach into three basis matrices corresponding to the three elementary polarimetry effects [6]

$$M \Leftarrow M_{Depol} M_R M_D = M_{Depol} M_{Pol} \qquad (S3)$$

Here, $M_{Depol}$, $M_R$ and $M_D$ describe the depolarizing effects, the retardance effect (phase difference between orthogonal polarizations), and the diattenuation effect (differential scattering for orthogonal polarizations) of the medium, respectively.

The most prominent polarization effect in our 1-D waveguided plasmonic samples was the linear retardance effect ($R_L$) (phase anisotropy). In addition, the samples also exhibited depolarization due to averaging effects in high NA microscopic geometry. The diattenuation effect was relatively weak. The linear retardance parameter $\delta$ was quantified from the elements of the non-depolarizing matrix $M_{Pol}$ as [2-5]

$$R_L = \cos^{-1}\left\{\sqrt{[M_{Pol}(2,2) + M_{Pol}(3,3)]^2 + [M_{Pol}(3,2) - M_{Pol}(2,3)]^2} - 1\right\} \qquad (S4)$$

The spectral variation of the Mueller matrix-derived linear retardance $\delta$ is shown in Figure S3. We have previously shown that the observed rapid variation of the $\delta$ parameter across the narrow resonance peak of the quasiguided mode ($E_0 = \hbar\omega_0 \sim 1.803\ eV$) is a characteristic signature of anisotropic Fano resonance in waveguided plasmonic crystal system. The $\delta$ parameter additionally possesses a frequency-independent background $\sim 0.25$ rad. The magnitude of $\delta$ at $E_0$ ($\omega_0$) was therefore used to decide the input elliptical polarization state ($S_{in} = [1\ -0.0493\ 0.9833\ -0.1752]^T$) for implementing the weak measurement schemes, as described in the manuscript.

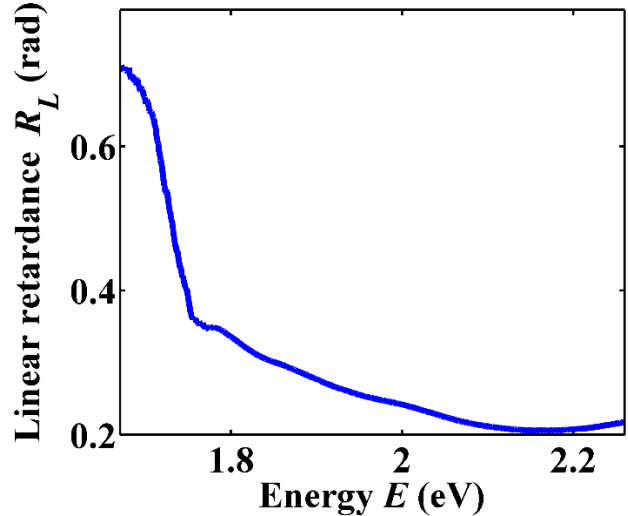

**Figure S3:** The spectral variation of the Mueller matrix-derived linear retardance $R_L$ parameter of the 1-D plasmonic crystal sample (corresponding to Mueller matrix $M$ of Figure 3a of the manuscript).

4. **Fabrication of plasmonic Structures**

The waveguided plasmonic crystal samples comprised of one dimensional (1-D) periodic gold (Au) grating on top of indium tin oxide (ITO) waveguiding layer coated on quartz substrate. We used electron beam lithography and metal deposition by thermal evaporation technique to fabricate these nanostructures (see Supporting information). First, the substrate was coated with two layers of e-beam resists of different sensitivity: Bottom layer was a copolymer based on methyl methacrylate and methacrylic acid (PMMA/MA 33 %), spin speed 6000 rpm and the top layer was Poly methyl methacrylate PMMA of molecular weight 950K, spin speed 8000 rpm. Both layers were baked on a hotplate at 175°C for 5 min immediately after spin coating. For e-beam exposure, we have used a Raith Elphy Quantum attachment on a Zeiss SIGMA Field Emission Scanning Electron Microscope (FESEM). We have used an accelerating voltage of 15KV and typical area dose of 600μC/cm$^2$. The exposed samples were developed in a solution of MIBK:IPA 1:3 for 60 s followed by iso-porpanol for 10s. We have thus created circular and elliptical holes in the e-beam exposed regions of the top layer resist of desired dimensions and a undercut in the bottom layer. Subsequently, thin layers of gold were deposited by thermal evaporation under high vacuum conditions, on the resist patterned samples and subsequently the resist was lift-off using acetone, leaving the desired pattern of Gold on the substrate. Here we report experiments on 1-D Au grating with dimension (width = 83 nm, height = 20 nm, centre to centre distance = 550 nm).

**References:**

1. Luk'yanchuk, B. *et al.* The Fano resonance in plasmonic nanostructures and metamaterials. *Nat. Mater.* **9**, 707-715 (2010).